# MID-INFRARED (MIR) OCT-based inspection in industry

N. P. García-de-la-Puente [1*], Rocío del Amor [1], Fernando García-Torres [1], Niels Møller Israelsen [3,4], Coraline Lapre [3], Christian Rosenberg Petersen [3,4], Ole Bang [3,4,5], Dominik Brouczek [2], Martin Schwentenwein [2], Kevin Neumann [6], Niels Benson [6], and Valery Naranjo [1]

[1] *Instituto Universitario de Investigación e Innovación en Tecnología Centrada en el Ser Humano, HUMAN-tech, Universitat Politècnica de València, Valencia, Spain.*
[2] *Lithoz GmbH, Vienna, Austria.*
[3] *Department of Electrical and Photonics Engineering, Technical University of Denmark, Kongens Lyngby, Denmark.*
[4] *NORBLIS ApS, Virum, Denmark.*
[5] *NKT Photonics, Blokken 84, 3460 Birkeroed, Denmark.*
[6] *airCode UG, Duisburg, Germany.*
*e-mail: napegar@upv.es

**Abstract**

This paper aims to evaluate mid-infrared (MIR) Optical Coherence Tomography (OCT) systems as a tool to penetrate different materials and detect sub-surface irregularities. This is useful for monitoring production processes, allowing Non-Destructive Inspection Techniques of great value to the industry. In this exploratory study, several acquisitions are made on composite and ceramics to know the capabilities of the system. In addition, it is assessed which preprocessing and AI-enhanced vision algorithms can be anomaly-detection methodologies capable of detecting abnormal zones in the analyzed objects. Limitations and criteria for the selection of optimal parameters will be discussed, as well as strengths and weaknesses will be highlighted.

## 1. Introduction

Minimising defects during the initial stages of industrial processes is crucial to prevent them from spreading to subsequent steps in production. Non-Destructive Testing (NDT) processes can provide high-quality data and identify defects [1]. The integration of artificial intelligence (AI) into these processes not only enables the early detection of potential defects but also allows for real-time monitoring of item progress throughout production stages. AI enhances the accuracy and speed of defect identification, contributing to the overall efficiency of quality control. Moreover, by leveraging AI, parameters can be continuously optimized, paving the way for enhanced quality in future manufacturing endeavours. The combination of NDT processes and AI presents a formidable approach to defect detection and prevention, offering a proactive strategy to ensure the highest levels of product quality by automatization and by circumventing human interpretation uncertainty factors [2].

In NDT, a range of imaging modalities is employed to assess the integrity of materials without causing damage. Radiographic techniques, such as X-ray and computed tomography (CT), provide detailed insights into internal structures. Ultrasonic testing utilises sound waves to detect flaws and assess material thickness. Thermography captures thermal patterns to identify irregularities in temperature distribution. Visual inspection, both direct and remote, remains a fundamental method for surface-level evaluations. Additionally, the growing prominence of Optical Coherence Tomography (OCT) introduces a high-resolution, non-invasive imaging approach.

OCT is an imaging technique where a light source is used to detect anomalies in both surface and sub-surface areas. This technology relies on interferometry to measure the time difference corresponding to the distances between the internal structures using a low-coherence near-infrared light beam (~800 nm or ~1300 nm) toward the material. The result is a high-quality image obtained without contact to the item under analysis. OCT has emerged in importance due to the wide variety of information it can provide: its high resolution, and its ability to gather complex 3-dimensional (3D) data [3]. Five years ago OCT based on mid-infrared light was realized providing deeper material imaging [4]. In recent years some research studies have applied machine learning and deep learning techniques to OCT anomaly detection problems. For example, Wolfgang et al. presented an evaluation method for OCT image analysis of pharmaceutical coatings based on deep convolutional neural networks [5]. In another study, Fin et al. applied a clustering approach (DBSCAN) to segment OCT images of pharmaceutical products, such as coated tablets, for real-time monitoring [6]. However, to the authors' knowledge, no previous work has focused on using AI techniques with OCT to detect defects in industrial manufacturing processes. In this paper, we propose a deep learning-based object detection model for defect detection in two industrial processes, wind turbine blade (WTB) generation and ceramics.

## 2. Related work

### 2.1 AI tools for NDT inspection in wind turbines

The development of new AI methods has led to the emergence of systematic, quantitative and automated tools for monitoring and diagnosing WTBs and making decisions within their life cycle. In this sense, Regent et al. developed methods using acoustics and efficient algorithms to detect WTB damage. These methods include the use of logistic regression and support vector machines for decision-making using binary classification algorithms [7]. Another study proposed a method using Gaussian Processes (GPs), which exploits the similarity between blades and the same environmental and operational variables to predict the edge frequencies of one blade based on the frequencies of another blade when they are in a healthy state [8]. Delamination in WTBs is a common structural issue that can lead to high costs. Early detection of delamination is essential to prevent breakages and downtime [9]. In [10], Jimenez et al. proposed a method for detecting delamination in WTB. The method used nonlinear autoregressive with an exogenous input (NARX) and linear autoregressive models to extract features from ultrasonic-guided waves, which are sensitive to delamination. Cracks are another common structural defect in WTB that can decrease the structure's lifespan. In [11], tree-based machine-learning algorithms were proposed to identify and detect cracks. The models were created by analysing the vibration response of the blade when excited by a piezoelectric accelerometer. Another study by Wang et al. found that irregular cracks can occur on blades before they break and used deep autoencoder (DA) models to predict the imminent failure of a blade using monitoring data [12].

### 2.2 ML tools for NDT inspection in ceramic

Non-destructive testing (NDT) allows the properties and integrity of ceramic parts to be examined without causing physical damage. Among the most commonly used NDT techniques for inspecting ceramic parts are ultrasonic inspection, infrared thermography and acoustic emission testing, among others [13]. The use of MIR-OCT technology on ceramic parts to detect defects, as on other types of objects, consists of examining changes in the refractive index of a sample. This non-destructive method does not require a contact medium, providing a 3D volume containing microscopic details of subsurface scattering. This system uses mid-infrared light with a center wavelength of 4 μm, which allows deep penetration into industrial alumina-based ceramics [14]. By scanning industrial ceramics with MIR-OCT, defects such as cracks, foreign particles and pores due to air bubbles inclusions present in the layer-to-layer transitions of the prints can be identified and correlated with the quality and performance of the printed component [15].

The combination of NDT and machine learning techniques offers several significant advantages in the inspection of ceramic parts. It enables more accurate and faster defect detection, which reduces inspection times and increases the efficiency of the production process. In addition, by integrating machine learning algorithms, inspection systems can adapt and continuously improve their ability to detect new types of defects or variations in part conditions. Some authors have used machine learning algorithms in combination with NDT techniques for the inspection of ceramic pieces, such as Support Vector Machine, Random Forest, and K-Nearest Neighbors algorithms on acoustic signals [16], [17]. Other authors have employed computer vision through convolutional neural networks for the analysis of images acquired using conventional and industrial photographic cameras, for the detection and characterisation of defects in ceramic pieces [18], [19], [20]. Regarding ultrasonic inspection, Naive Bayes classifiers, KNNs and Long Short-Term Memory (LSTM) networks have been used for detecting flaws [21].

## 3. Material and Methodology

The use case scenario dataset consists of three volumes of coated glass-fibre composite from Sample B in [22] and three volumes in ceramic green state (lithography-based parts). As can be seen in Figure 1, two types of defects (voids and cracks) are considered in the fibre volumes and three types (voids, surface irregularities and agglomerates) in the ceramic ones.

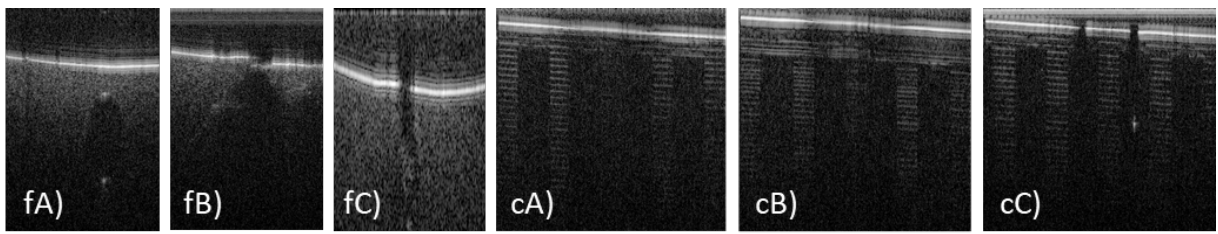

**Figure 1**. Slice samples by volume. A void can be seen in fA and cracks in fB and fC. Four voids can be seen in the ceramic samples, as well as surface irregularities and a shiny agglomerate in the slice shown from cC.

Volume slices have 400 x 400 dimensions (except Fiber C, which is 200 x 400) and are labelled in a supervised manner using squared Bounding Boxes that enable the anomaly location. The volumes are very varied in the number of defects, which produces a clear imbalance in certain classes (Table 1). To address the difficulty of detection in such a limited dataset, the Leave-One-Out Methodology is employed, where two volumes of material are used for training and one for validation.

**Table 1.** Number of Bounding Boxes labelled per volume.

| Volume | Voids | Cracks | Surface irregularities | Agglomerates |
|---|---|---|---|---|
| Fiber A (fA) | 87 | 78 | - | - |
| Fiber B (fB) | 29 | 248 | - | - |
| Fiber C (fC) | - | 31 | - | - |
| Ceramic A (cA) | 459 | - | 99 | 31 |
| Ceramic B (cB) | 270 | - | 16 | 3 |
| Ceramic C (cC) | 335 | - | 174 | 5 |

Because of its effectiveness and efficiency [23], YOLOv8n was selected as a model suited to the application in industrial settings for defect detection. It divides the image into a grid and predicts Bounding Boxes and class probabilities for each grid cell. The training hyperparameters selected for the current study determine a batch size of 16 images, 200 epochs, considering that the training stops if there is no improvement in the last 50 epochs. At the hardware level, the resources that have been allocated for training correspond to NVIDIA RTX 3090 24 GB x1, 525.60.11 drivers & CUDA 12.0, MSI Z270 (MS-7A63); 32 GB and Intel i7-7700K (4.2 GHz).

## 4. Results

The key metrics in the evaluation of trained object detection models are the precision and the inference time in which an image is predicted. Mean Average Precision (mAP) allows the measurement of the prediction accuracy and Intersection over Union (IoU) to measure the overlap between two Bounding Boxes. An IoU limit of over 50% is defined to determine whether a prediction is regarded as true. For the presented use case, maximum mAP50 metrics achieved during inference were collected and summarized (Table 2).

**Table 2.** Calculated mAP50 per class, training and inference time.

| Model YOLOv8n | Voids | Cracks | Surface irreg. | Agglom. | Train/Inference time |
|---|---|---|---|---|---|
| train fB fC - inference fA | 0.238 | 0.386 | - | - | 0.136h/1.0ms |
| train fA fC - inference fB | 0.785 | 0.103 | - | - | 0.128h/0.9ms |
| train fA fB - inference fC | - | 0.0351 | - | - | 0.181h/1.7ms |
| train cB cC - inference cA | 0.701 | - | 0 | 0.501 | 0.134h/2.1ms |
| train cA cC - inference cB | 0.56 | - | 0.141 | 0.233 | 0.099h/0.9ms |
| train cA cB - inference cC | 0.792 | - | 0.273 | 0.284 | 0.125h/1.4ms |

The results from the YOLOv8n model trained with the dataset of coated glass-fibre composite shows that the void defect in Fiber B volume is easily detected. However, in general, it cannot correctly identify cracks due to the small dataset used. In parallel, the YOLOv8n model trained with the ceramic data performs well for the case of voids but not so well for surface irrregularities and agglomerates. Some examples of inferences with correctly detected defects are shown in Figure 2.

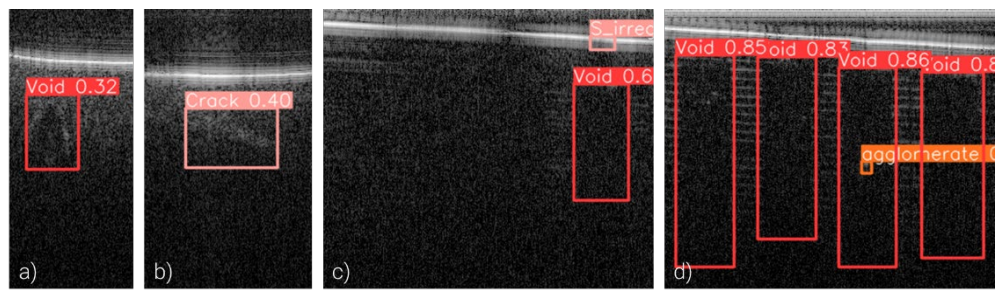

**Figure 2**. Samples of defect detection. Void and crack detected in fB (a and b), Surface irregularity and void detected in cB (c), Voids and agglomerate detected in cC (d).

Finally, the interest of using such lightweight models for their speed is to be noted. This can be seen in the last column of Table 2, where the training sessions last ten minutes or less and the inferences are around a millisecond in duration.

## 5. Conclusion

In conclusion, the application of YOLO (You Only Look Once) for defect detection in OCT images of turbine blades and ceramics has demonstrated remarkable efficacy, particularly in identifying cracks and voids. This study represents a preliminary exploration, constrained by a relatively limited volume of available images. Future research will focus on acquiring and incorporating a larger dataset. A larger and more diverse dataset would likely improve the generalizability of the model, further enhancing its ability to detect defects.

## Acknowledgements

This work has received funding from Horizon Europe, the European Union's Framework Programme for Research and Innovation, under Grant Agreement No. 101058054 (TURBO) and No. 101057404 (ZDZW). Views and opinions expressed are, however, those of the author(s) only and do not necessarily reflect those of the European Union. Neither the European Union nor the granting authority can be held responsible for them. The work of Rocío del Amor has been supported by the Spanish Ministry of Universities (FPU20/05263). This work has received funding from Spanish Ministry of Science and Innovation for the project ASSIST (PID2022-140189OB-C21).